\documentclass[cryptography,article,accept,pdftex,moreauthors]{Definitions/mdpi} 
\firstpage{1} 
\pubvolume{1}
\issuenum{1}
\articlenumber{0}
\pubyear{2023}
\copyrightyear{2023}
\externaleditor{Academic Editor: Firstname Lastname}
\datereceived{17 August 2023 } 
\daterevised{ 9 November 2023} 
\dateaccepted{13 November 2023 } 
\datepublished{date } 
\hreflink{https://doi.org/} 
\Title{Secure Instruction and Data-Level Information Flow Tracking Model for RISC-V}

\TitleCitation{Secure Instruction and Data-Level Information Flow Tracking Model for RISC-V}

\Author{Geraldine Shirley Nicholas *
 $^{\dagger}$, Dhruvakumar Vikas Aklekar, Bhavin Thakar $^{\dagger}$ and Fareena Saqib *}

\AuthorNames{Geraldine Shirley Nicholas, Dhruvakumar Vikas Aklekar, Bhavin Thakar and Fareena Saqib}

\AuthorCitation{\textls[-20]{Nicholas, G.S.; Aklekar, D.V.;} Thakar, B.; Saqib, F.}

\address[1]{%
{Electrical and Computer Engineering, University of North Carolina at Charlotte, Charlotte, NC 28262, USA}; daklekar@uncc.edu (D.V.A.); bthakar1@uncc.edu (B.T.)}

\corres{\hangafter=1 \hangindent=1.05em \hspace{-0.82em}{Correspondence: gnichola@uncc.edu (G.S.N.); fsaqib@uncc.edu (F.S.)}}

\firstnote{\hangafter=1 \hangindent=1.05em \hspace{-0.82em}{These authors contributed equally to this work.}}

\abstract{With the proliferation of electronic devices, third-party intellectual property (3PIP) integration in the supply chain of the semiconductor industry and untrusted actors/fields have raised hardware security concerns that enable potential attacks, such as unauthorized access to data, fault injection and privacy invasion. Different security techniques have been proposed to provide resilience to secure devices from potential vulnerabilities; however, no one technique can be applied as an overarching solution. We propose an integrated Information Flow Tracking (IFT) technique to enable runtime security to protect system integrity by tracking the flow of data from untrusted communication channels. Existing hardware-based IFT schemes are either fine-, which are resource-intensive, or coarse-grained models, which have minimal precision logic, providing either control-flow or data-flow integrity. No current security model provides multi-granularity due to the difficulty in balancing both the flexibility and hardware overheads at the same time. This study proposes a multi-level granularity IFT model that integrates a hardware-based IFT technique with a gate-level-based IFT (GLIFT) technique, along with flexibility, for better precision and assessments. Translation from the instruction level to the data level is based on module instantiation with security-critical data for accurate information flow behaviors without any false conservative flows. A simulation-based IFT model is demonstrated, which translates the architecture-specific extensions into a compiler-specific simulation model with toolchain extensions for Reduced Instruction Set Architecture (RISC-V) to verify the security extensions. This approach provides better precision logic by enhancing the tagged mechanism with 1-bit tags and implementing an optimized shadow logic that eliminates the area overhead by tracking the data for only security-critical modules.}


\begin{document}


\section{Introduction}

The rise in smart technology, Internet of Things (IoT) and Heterogeneous Systems on Chips (SoCs) have facilitated the connected ecosystem; however, automation has introduced several supply chain and runtime attack surfaces. Securing connected interfaces and tracking the authenticity of data from untrusted communication channels is a major security concern. Memory corruption vulnerabilities, code injection, buffer overflow attacks and other software-based attacks compromise untrusted channels to control the flow of applications with malicious data.

Security measures to protect interfaces and data propagation via channels are critical in the framework of preserving the confidentiality and integrity of data. The IFT approach is used for the detection of data attacks and memory corruption by tracking the flow of data through the untrusted communication channels~\cite{ref-p1,ref-p2}. The IFT approach labels malicious data with tags to denote security policies and tracks data propagation to determine the authenticity of data at runtime. IFT features may be used as static verification during the design phase and provide dynamic checking during functional modes at runtime.

The IFT model is classified on the level of abstraction, precision logic and verification techniques~\cite{ref-p3}. This study proposes a novel technique that supports both instruction-level and gate-level secure propagation. The existing techniques include fine-grained models that use high-precision data-dependent labels and coarse-grained models with upper-bound data labels. These individual models contribute to a particular granularity-based precision logic, resulting in limited security properties supported via the given model. Gate-level information flow tracking is based on shadow logic supporting precise models but results in complex design computations, leading to scaling issues and complexity in GLIFT logic. On the contrary, Dynamic Information Flow Tracking (DIFT)~\cite{ref-p4} is an Instruction Set Architecture (ISA)-based hardware approach with conservative tracking rules. This approach provides coarse-grained granularity but lacks adaptability for different ISAs, thus limiting its applications.

Most DIFT designs focus on protecting the target software but fail to protect the information flow between third-party IPs (3PIPs), which can perform security-critical functions~\cite{ref-p5}. Furthermore, SoC-level-based DIFT designs with bus extensions for 3PIP protection are not flexible, leading to system overheads. Therefore, by considering the precision and granularity of a system, an IFT model may provide both levels of granularity for better precision, along with securing the integrity of the system.
The proposed approach makes use of instruction-based and gate-level-based models, providing both fine- and coarse-grained granularity by integrating the models and optimizing systems to enhance the performance of the combined models. This approach further implements a novel hardware-to-software simulation-based IFT model to exhibit a correlation with toolchain modification and leverages the RISC-V platform to provide security extensions for the processor ISA.

Hardware-based IFT models often modify architectures or use co-processors, leading to overhead and proprietary complexity. Therefore, an open-source, supporting, modular and extensible RISC-V base platform is used to support secure applications. The proposed technique leverages the architecture’s flexibility by adding secure extensions and the IFT model with RISC-V ISA. Gate-level IFT is applied to security-critical datapath modules with enhanced shadow logic to minimize the design complexity of the model.
This study makes the following contributions:
\begin{itemize}
\item	A RISC-V-based coarse-grained hardware-based Information Flow Tracking framework with a tagged mechanism is proposed and demonstrated at an architectural level, which detects security violations at runtime. 
\item	A RISC-V-based fine-grained gate-level-based Information Flow Tracking framework for a security-critical datapath using optimized shadow logic is demonstrated with minimal overhead.
\item	Toolchain extensions with a hardware-to-simulation-based correlation IFT RISC-V model are implemented and verified with better precision.
\end{itemize}

The paper is organized as follows: Section~\ref{sec2} describes the background work of the existing IFT models with a RISC-V architecture. Section~\ref{sec3} illustrates the threat model, and Section~\ref{sec4} presents the proposed design approach. The experimental evaluation and results of the proposed model are described in Section~\ref{sec5}. Security analysis, limitations and future work are discussed in Sections~\ref{sec6}~and~\ref{sec7}. 
\section{Background Studies}\label{sec2}

Information flow tracking techniques support hardware- and software-based implementations and depend on the explicit and implicit flow of data to design the conditional behaviors for the model. The following section discusses the existing IFT approaches based on precision granularity models.

\subsection{Coarse-Grained IFT Models}

Architectural-level IFT implementations provide a coarse-grained precision logic with a dedicated tag mechanism for security policies and metadata propagation. Hardware-based approaches depend on the architectural features and datapath of an ISA model. Coarse-grained models track control sensitive information in an application, along with program variables and other independent labels. Access and information flow with tags can be associated with a dedicated co-processor or a modified ISA with tag modules. IFT models protect systems from buffer overflow attacks and memory corruption via identifying malicious data~\cite{ref-p6,ref-p7}. 

FlexiTaint is an architectural-level IFT model~\cite{ref-p8} that supports an accelerator with tainted security policies and extends the processor’s datapath for tag propagation. DIFT~\cite{ref-p9} is a hardware-based approach with an ARM co-processor to track and debug traces using static analysis. Hardware-Assisted Data-Flow Isolation (HDFI)~\cite{ref-p10} and HyperFlow~\cite{ref-p11} are hardware-based RISC-V implementations that implement tagged mechanisms and security policies for information flow control. \textls[-20]{Exploitable Buffer Overflow Detection by Information Flow Tracking (BOFT)~\cite{ref-p12} is an automated framework with an extensive library for explicit and implicit IFT that integrates formal verification with IFT and leverages symbolic execution but adds extensive instrumentation for taint propagation. Mixed-mode IFT (MIT)~\cite{ref-p13} provides byte granularity by decoupling the tracking logic from program execution by using the taint semantics at compile time, which is dependent on runtime logs. }

Christian Palmiero et al.~\cite{ref-p14} proposes a feasible RISC-V core without any runtime overhead to detect memory corruption attacks by adding security tags to user-supplied inputs. Self-authenticated~\cite{ref-p15} secure boot with application runtime security provides information flow tracking with shared memory to reduce the area overhead of the system. Hybrid-mode IFT with Taint (HIT) semantics extraction and replay~\cite{ref-p16} integrates static information analysis and DIFT via compile-time extraction to detect sensitive data leakage with a modest performance overhead.
Tagged Memory Assisted for Fine-grained Data-Flow Integrity (TMDFI) models~\cite{ref-p17} uses lowRISC open-source tag-memory-based architecture for runtime data integrity with a limitation of an 8-bit memory tag that can support 256 tags sufficient for small embedded applications. Tagged memory based on hardware uses static analysis techniques to perform security policies on target programs to mitigate data-oriented attacks~\cite{ref-p18}. 

FineDIFT~\cite{ref-p19} is a hardware-based mechanism that generates data-flow graphs of a running process with a co-processor to provide flexibility, but the metadata storage requirements result in several limitations of software instrumentation.  Kejun Chen et al. summarize current DIFT solutions, referring to the over-tainting problem that leads to high false positive rates~\cite{ref-p5}. Security policies with customization can be achieved by adopting parallel tag propagation schemes in these models. Optimal-decision-based DIFT~\cite{ref-p20} uses an analytical algorithm for indirect flow propagation with an arbitrary number of tag types for flexibility. Cryptographic Return Address Stack (CRAS)~\cite{ref-p21} and Return-Address-Stack-Based Side-Channel LEakage(RASSLE)~\cite{ref-p22} are return address stack models that protect the stack from leakage channels. A tag-based model results in architectural overhead, as the memory is modified to incorporate tag bits or complex lattice structures or built for securing the data. An efficient model needs to limit the hardware design overheads and provide IFT capabilities in the hardware design.  

\begin{figure}[H]
\includegraphics[width=8.5 cm]{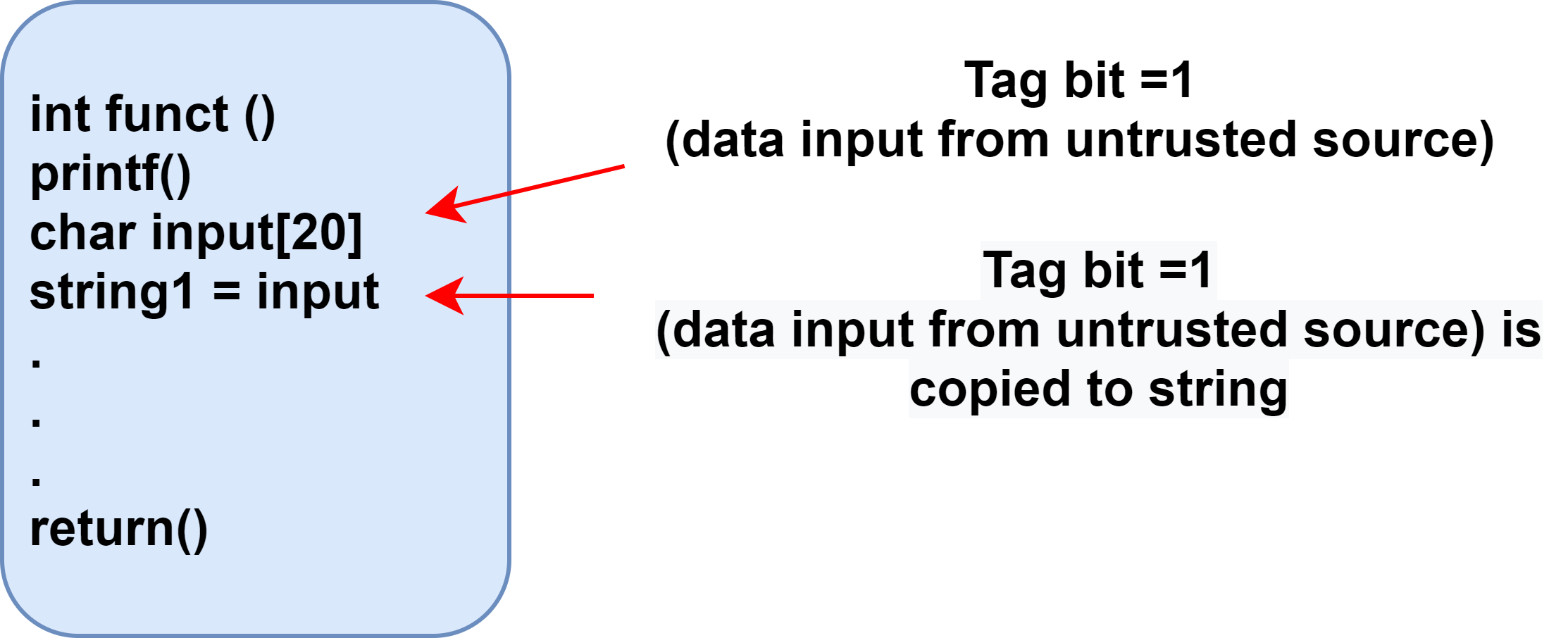}
\caption{Tag bits assigned to untrusted sources via the IFT module.\label{fig1}}
\end{figure}

A tagging mechanism assigns a tag bit to untrusted data and tracks the propagation of data using tag bits. Figure~\ref{fig1} illustrates an example in which incoming data from an untrusted source (input [20]) is assigned a tag bit, and the IFT module in the background tracks the data using the value of the tag bits to indicate if the data are spurious or safe.
All incoming data are assigned a tag bit, with a value of 1 or 0 indicating unsafe or trusted data, respectively. If the data are copied to string1, it is identified as an untrusted source, and the IFT module tracks the data associated with string1. This approach is based on hardware design implementation wherein a one-bit tag is assigned to untrusted sources and an IFT module, which consists of different units, tracks the tag bits to authenticate the data based on the security policies for the information flow. 
 
\subsection{Fine-Grained IFT Models}

IFT implementations also support fine-grained models with data labels associated with only data flow and do not focus on control-flow transfers. This approach focuses on a gate-level IFT model, which is a fine-grained model associated with Boolean gates in which the information flow appears at the gate-level netlist and provides better precision logic. GLIFT~\cite{ref-p23} is a shadow-logic-based technique that adds logic to all the gates, resulting in design complexity and overhead. \textls[-20]{Data-flow logic~\cite{ref-p24} serves as a backbone, which is used in several proposed techniques, including~\cite{ref-p25,ref-p26,ref-p27}, where optimized labeling and enhanced encoding techniques help to reduce complexity but affect the precision logic of the system. An asset-based GLIFT~\cite{ref-p28} model provides structural checking with security properties. }

Gate-level-based leakage detection~\cite{ref-p29} with parser and logic modules for formal verification detects leaky paths but results in intense computation complexity for large designs. A multi-bit label tracking model~\cite{ref-p30} quantitatively detects information leakage with area constraints, as the multi-bit labels are directly proportional to the number of gates in the circuit. A unified model for gate-level propagation~\cite{ref-p31} generates synthesizable propagation logic to be used in Electronic Design Automation (EDA) tools where attribute labels at different levels of precision are addressed for the faults, with the flexibility to be used in different emulation platforms.

\begin{figure}[H]

\includegraphics[width=8.5 cm]{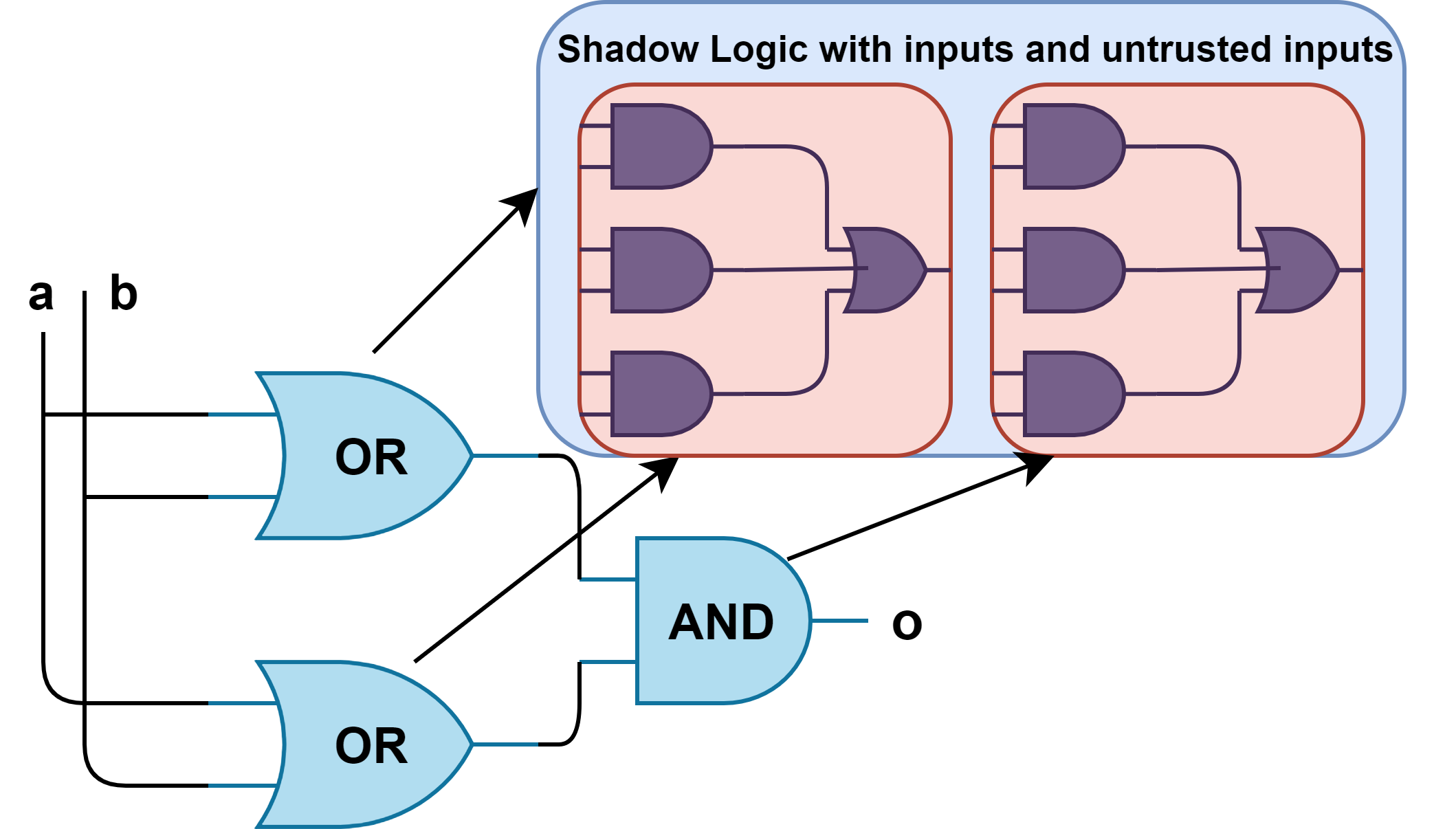}
\caption{Gate-level IFT with shadow logic.\label{fig2}}
\end{figure}

A gate-level-based IFT model consists of gates and shadow logic for all gates, as represented in Figure~\ref{fig2}. For each gate, a separate shadow logic is implemented (a and b are the inputs with an output o) and the shadow consists of inputs, along with untrusted inputs, which affect the output. Such a scheme results in additional area overhead. The proposed scheme reduces the complexity of the gate-level model and enables tracking of the data-critical modules, such as crypto engines and security accelerators, rather than tracking the whole module.  This proposed technique integrates the tagged mechanism and provides coarse and fine granularity in tracking the data.

\subsection{Modular ISA (RISC-V) }

RISC-V is an open-source, extensible ISA with a minimal instruction set, providing a flexible customized processor. RISC-V supports several security applications, such as Common Evaluation Platform (CEP)~\cite{ref-p32}, to identify security properties and, thus, stands as a major concern in the security of the underlying design model. The RISC-V security committee has proposed an abstraction-augmented aISA that extends a bridge between hardware and software beyond the traditional ISA for control~\cite{ref-p33}. RISC-V extensions to support the IFT model provide strong timing-sensitive security behavior by leveraging the flexibility of the architecture, which is otherwise not possible with traditional ISAs.

The RISC-V toolchain consists of GNU Compiler Collection (GCC) and Clang/Low-Level Virtual Machine (LLVM) compilers and provides different backend library support for customization~\cite{ref-p34}. Simulator models, such as Spike and Qemu, are used as reference models for RISC-V ISA~\cite{ref-p35,ref-p36}. Isadora~\cite{ref-p37} framework uses a minimal testbench to run an automated trace generator for information tracking, which limits access to the full RISC-V toolchain, leading to false positives. The toolchain modification with customization in a simulation model achieves a correlation between the hardware and software and can enable the evaluation of security extensions effectively. The proposed work integrates hardware and software by translating the architecture-specific extensions to compiler-specific simulation models, thereby enabling system-level security with higher accuracy.

\section{Threat Model}\label{sec3}

The threat model assumes the device is in an untrusted field and has the runtime vulnerability of stealthy modifications to the input channels to circumvent the security layer. An adversary can use an untrusted communication channel to exploit vulnerable codes by providing malicious inputs to the system. The application is further vulnerable to permitting modification in the configuration parameters through software-based attacks. This work focuses on memory corruption through buffer overflow attacks and return address attacks. Furthermore, the proposed techniques evaluate data-level leakage detection in the functional modules.

\subsection{Buffer Overflow Attacks }

The memory structure is an important aspect of an architecture in which the stack plays an important role in storing the temporary variables created via a function. When a program is executed, the information is stored, along with the parameters, return address and base pointer. An adversary can exploit user inputs without bound checking with excess data that can overwrite the return address~\cite{ref-p38} and other memory locations~\cite{ref-p39}, with malicious data leading to system compromise or running malicious codes that may leak critical information from the system.

\begin{figure}[H]

\includegraphics[width=6.5 cm]{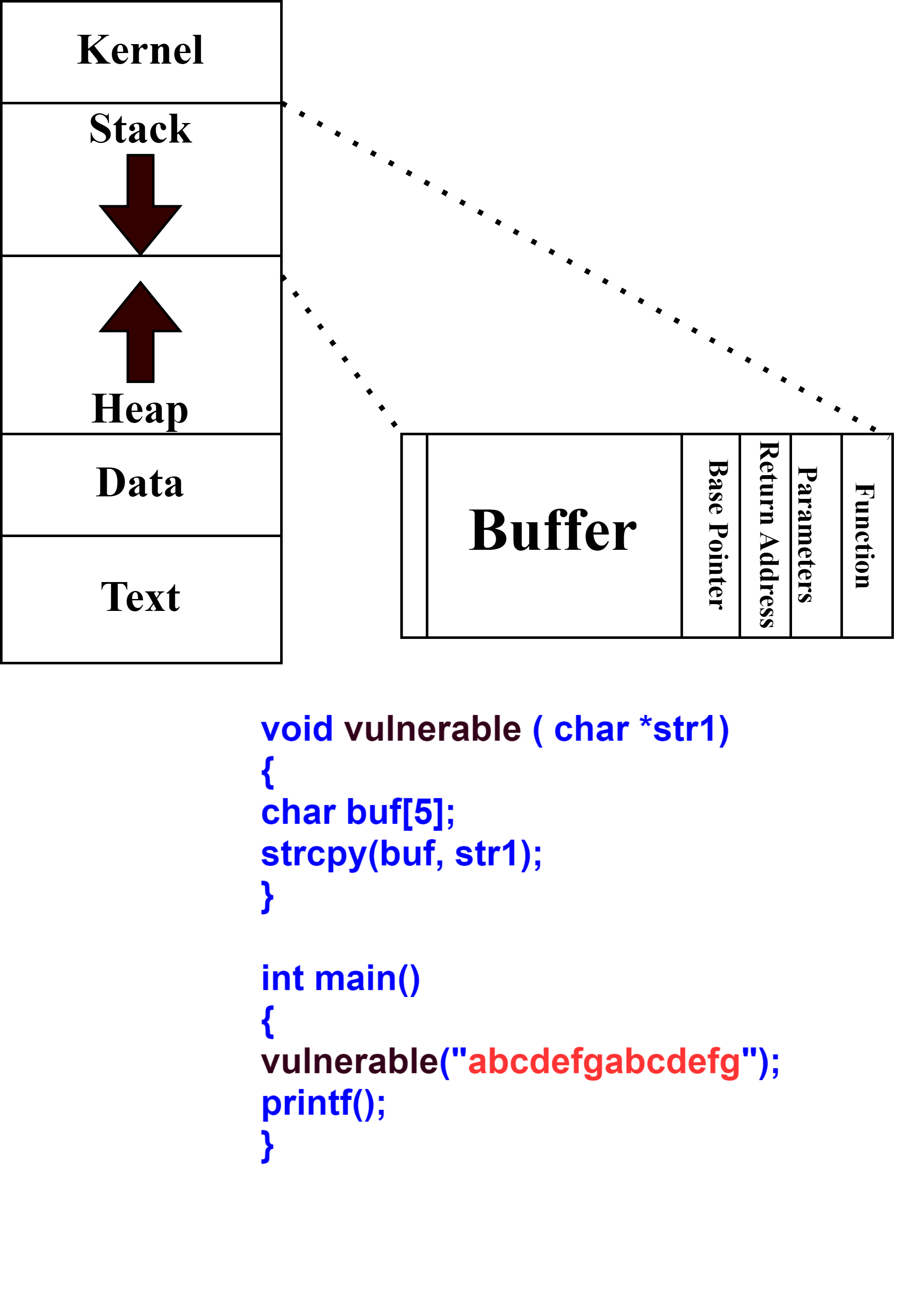}
\caption{Memory structure with buffer overflow attack.\label{fig3}}
\end{figure}

Figure~\ref{fig3} shows the overall memory structure with a buffer overflow attack caused by the sample program. The memory structure consists of the stack and heap structure along with other sections to be stored in the memory. In Figure~\ref{fig3}, a string (str1) is passed with a size larger than the buffer size, causing the return address to be overwritten. This example shows the buffer overflow attack and the need for security policies to detect and eliminate this scenario before the return address is modified.
A security policy can assign tag bits to the data return address to catch the attack and prevent corrupting the parameter from a buffer overflow attack.

\subsection{Return Address Attacks}

The return address attack targets return to libc, in which an adversary injects a malicious code that executes a hidden function in the background, which can go undetected by the system~\cite{ref-p40}.  This attack does not need an executable stack or a shellcode but instead causes the main code to jump to the malicious program.

\begin{figure}[H]

\includegraphics[width=8.5 cm]{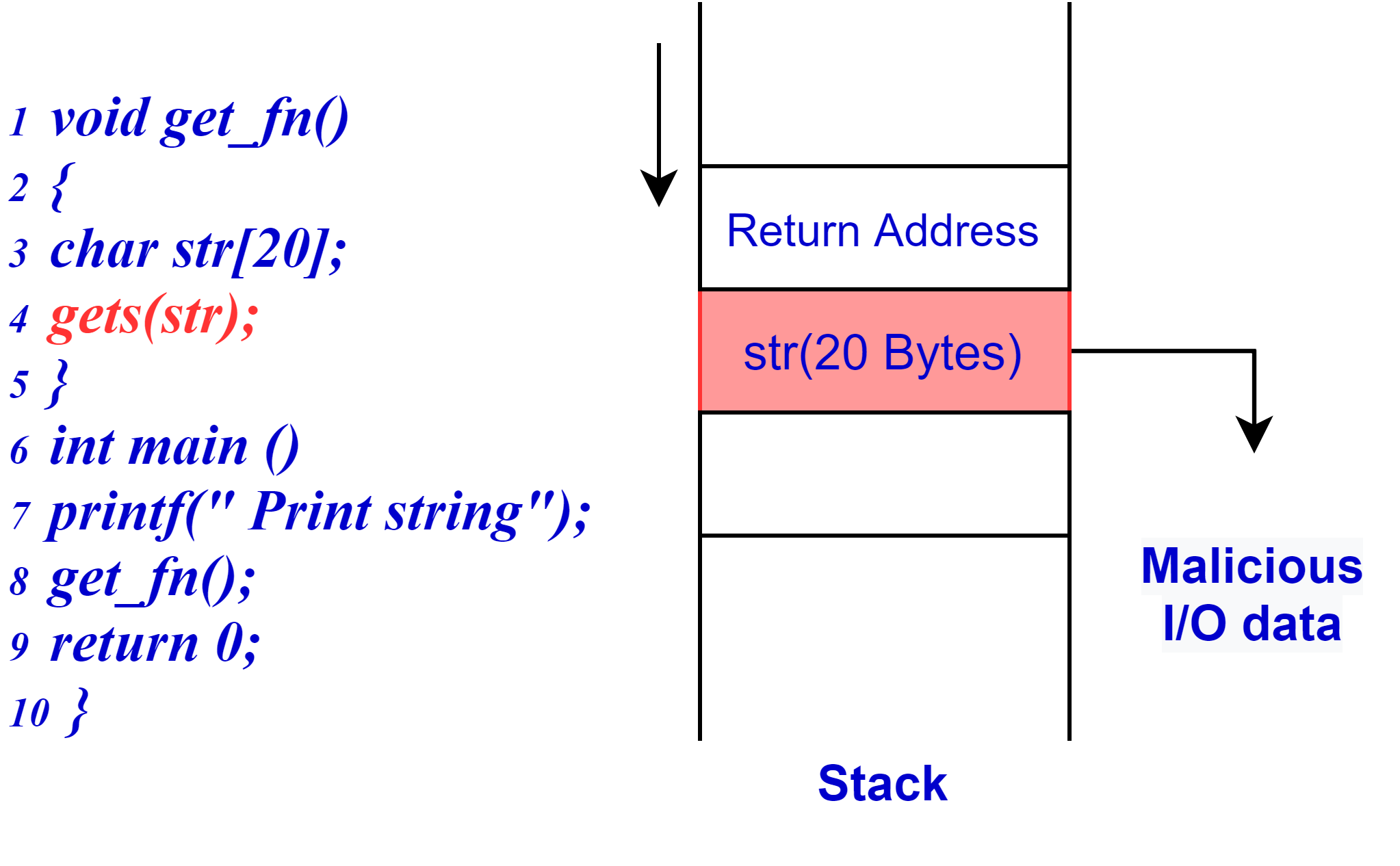}
\caption{Buffer return address attack.\label{fig4}}
\end{figure}

Figure~\ref{fig4} illustrates the return address attack via a procedure call where the input data from an untrusted communication channel affects the return address. In this example, the return address of the main function is saved in the stack, and it calls the get\_fn() for which the local variables are created and specific space is allocated to the buffer. The input/output string data from the untrusted communication channel exceeds the limit of the buffer capacity and overwrites the return address. The return address is now modified via the procedure call function, which assigns the return address pointing to the adversary’s malicious code to be executed. 

To protect the return address from such attacks, an address space randomization protection scheme is used in most systems but this technique is vulnerable to side-channel attacks. Hence, to protect the system from return address attacks, new custom instructions can be implemented to track the flow of the address using tags in an IFT model.

\subsection{Data-Level Leakage}

Hardware designs are potentially vulnerable to the leakage of sensitive information from security-critical modules, such as cryptographic engines and data processing accelerators, where maliciously added hardware Trojan can be activated to trigger a payload condition~\cite{ref-p41}. The analysis or tracking of data requires access to a gate-level netlist to check the inputs that influence the outputs. Modeling the integrity properties of data at the gate level is performed via gate-level IFT using shadow logic~\cite{ref-p23,ref-p42}.

\begin{figure}[H]

\includegraphics[width=10.5 cm]{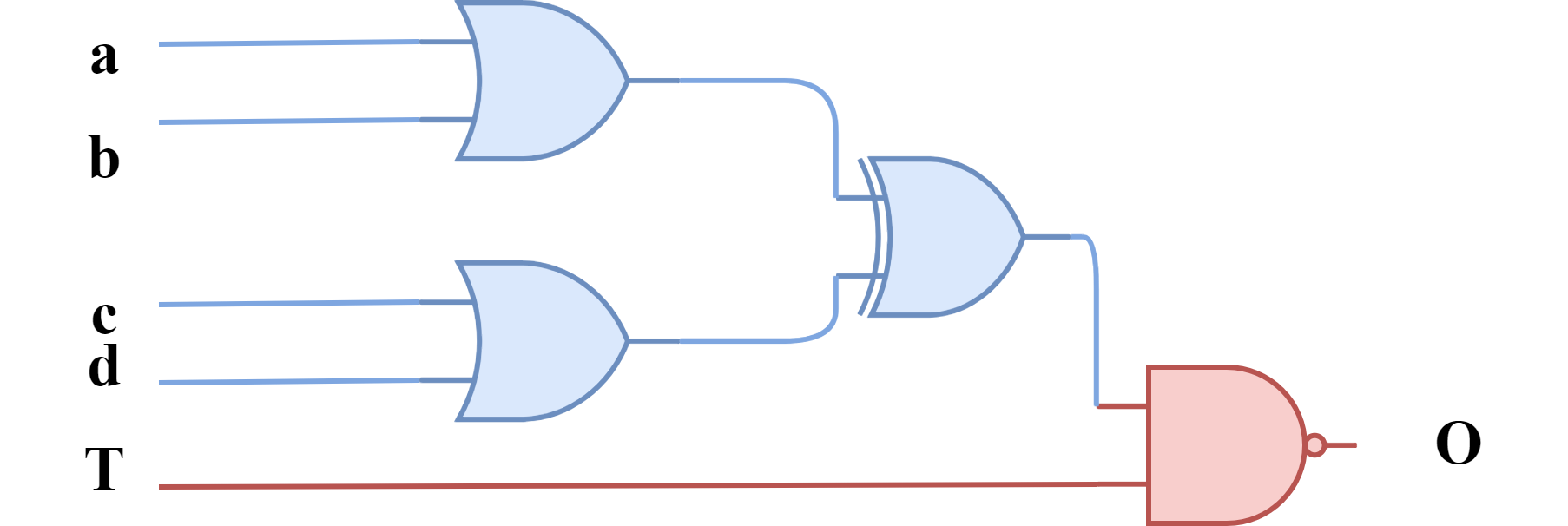}
\caption{Trojan insertion in gate-level circuits.\label{fig5}}
\end{figure}

Figure~\ref{fig5} shows an example of a Trojan insertion in a gate-level circuit consisting of logic gates. The inputs (a, b, c, and d) are given to OR gates, and the result is XORed, which, in turn, is given as input to the NAND gate. A Trojan input (T) is fed as another input to the NAND gate, which acts as a hidden trigger. When the trigger input is 1, the output is affected via the input, resulting in an activated signal that triggers the hidden function to be enabled. The secret keys are leaked or replaced by the function executed by an adversary. Thus, the violation of key integrity is tracked by information flow properties, which are implemented using gate-level IFT models providing fine-grained precision logic. 

\section{CF-IFT Technique---Integrated Coars- and Fine-Grained IFT}\label{sec4}

The proposed 
hardware-based IFT technique is an integrated coarse-grained-based data tracking technique enriched with a fine-grained modular gate-level IFT module for security-critical datapaths. The proposed scheme is demonstrated on a simulation model with toolchain support to track return addresses to prevent memory corruption and software attacks. To the author’s knowledge, this is a novel effort that integrates toolchain support for RISC-V security extensions along with multi-granularity for better accuracy and preventing false positives in the system. 

\subsection{Architectural Model}

The architectural model is demonstrated on a RISC-V platform enriched with security extensions in the model, which detects and stops software attacks on memory to mitigate corruption-based attacks. Security policies protect a system from different untrusted sources; these countermeasures are demonstrated by protecting the return address with an extendable tag mechanism. The tag mechanism module consists of a modular approach that assigns tag bits and tracks the propagation of the tags during runtime to detect spurious data.

\subsubsection{RISC-V}

A RISC-V core features both 32- and 64-bit variants with fixed-length 32-bit instructions and variable-length encoding for customizable applications~\cite{ref-p43}. 
The proposed scheme integrates the security features and extends them to the instruction set level. RISC-V is a load--store architecture, in which only load and store instructions access the memory; the IFT model focuses on the load and store instructions to check the integrity level of the memory contents. The loads are encoded in I-type format, and the store instructions are encoded in S-type format. Two new instructions are added to the ISA, which is used to provide security checks for the 1-bit tag in the tag mechanism~\cite{ref-p44}. Based on the load and store encoding specified in the RISC-V core, the new instructions are as follows:

\begin{itemize}
\item   In the load instruction encoding: LDTCHECK;
\item	In the store instruction encoding: SDTCHECK.
\end{itemize}

\textls[-20]{The new store instruction encoding is used to store data and assign a tag bit to the address of the data, that is, if the data are evaluated as spurious, a tag bit is assigned as 1 or 0. The new load instruction encoding is used to load the data and check if the tag bit is 1 or 0. A tag module is designed, in which an array table is assigned for the tag bits. The array table is assigned a separate virtual access with a tag cache instead of the main memory to protect the table access. The Control and Status Register (CSR) in the ISA offers custom address space for unused addresses, which is used to assign a new status for the tag values and mismatch conditions, which results in an exception. The core is extended to add the tag bit along with the read and write requests to accommodate the new tag bit. The tag module is used to check the address of the load and store instructions to reduce the overhead in the architecture.}

\subsubsection{Tag Mechanism}

The tag module consists of the tag initialization, tag propagation and tag checking modules. The usage of 1-bit tags reduces the overhead, and additionally, memory access takes place separately in the tag cache. The tag initialization module assigns tags to all the data addresses from untrusted sources using the custom-added load and store instructions. The purpose of the tag propagation unit is to track the tag bits from all the data addresses and store the details of the tags in the tag cache. Minimal information along the path is stored in the tag cache to avoid overhead. Finally, the tag-checking module is used to check the tag bits for all the data addresses after a procedure call, and if there is a mismatch in the tag bit, then an exception is raised. 

\begin{figure}[H]

\includegraphics[width=9.5 cm]{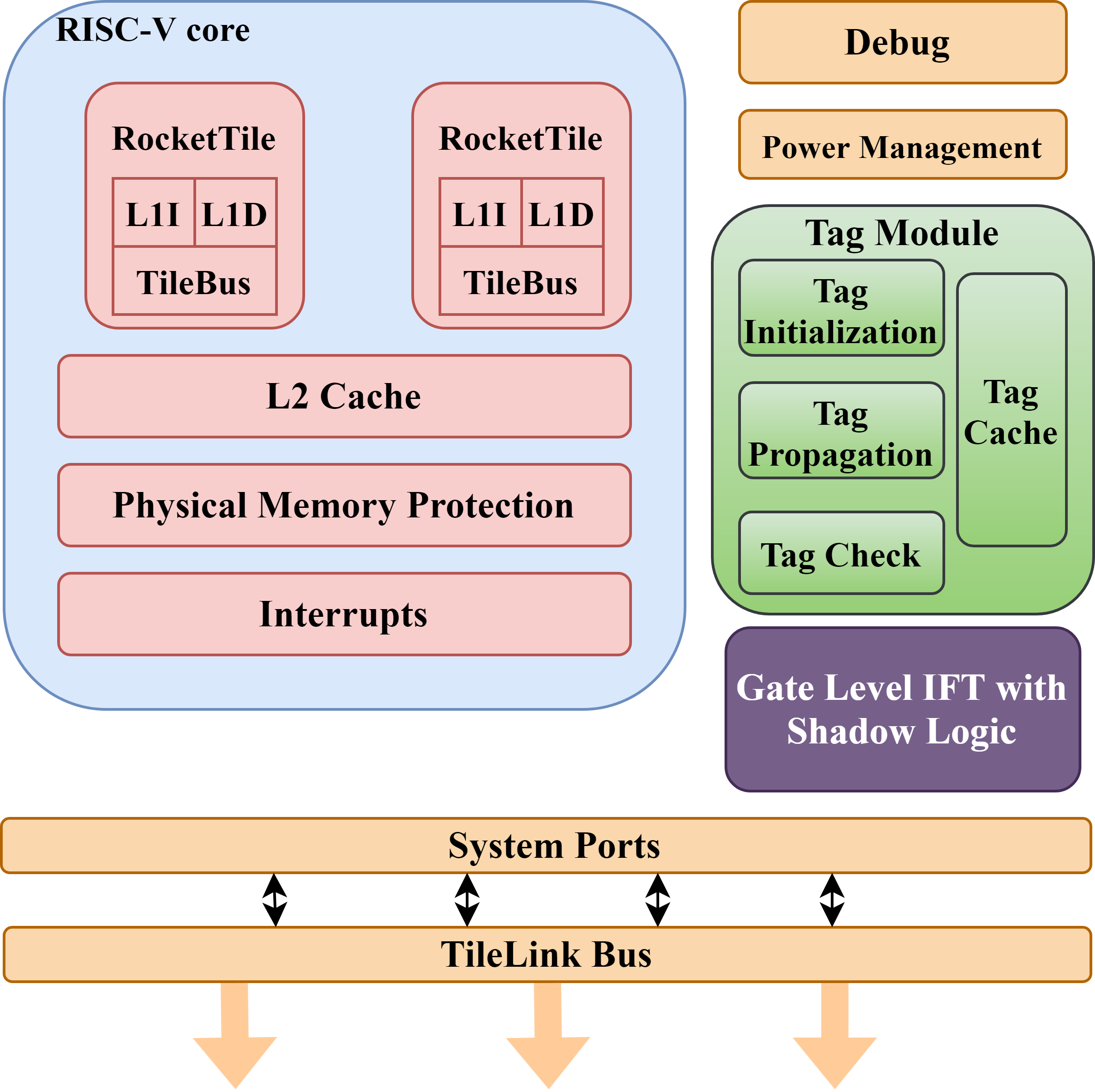}
\caption{RISC-V core with tag mechanism and gate-level IFT.\label{fig6}}
\end{figure}

The checking unit checks all the properties associated with the security policies for the data. Based on the security policies, the data are assigned as spurious and tracked completely to protect them from being modified or affecting the memory, leading to different attacks. Figure~\ref{fig6} shows the overall system design with the RISC-V core Tag mechanism and gate-level IFT with shadow logic, in which individual units are used to protect the memory from software attacks and data leakage.


\begin{figure}[H]

\includegraphics[width=8.5 cm]{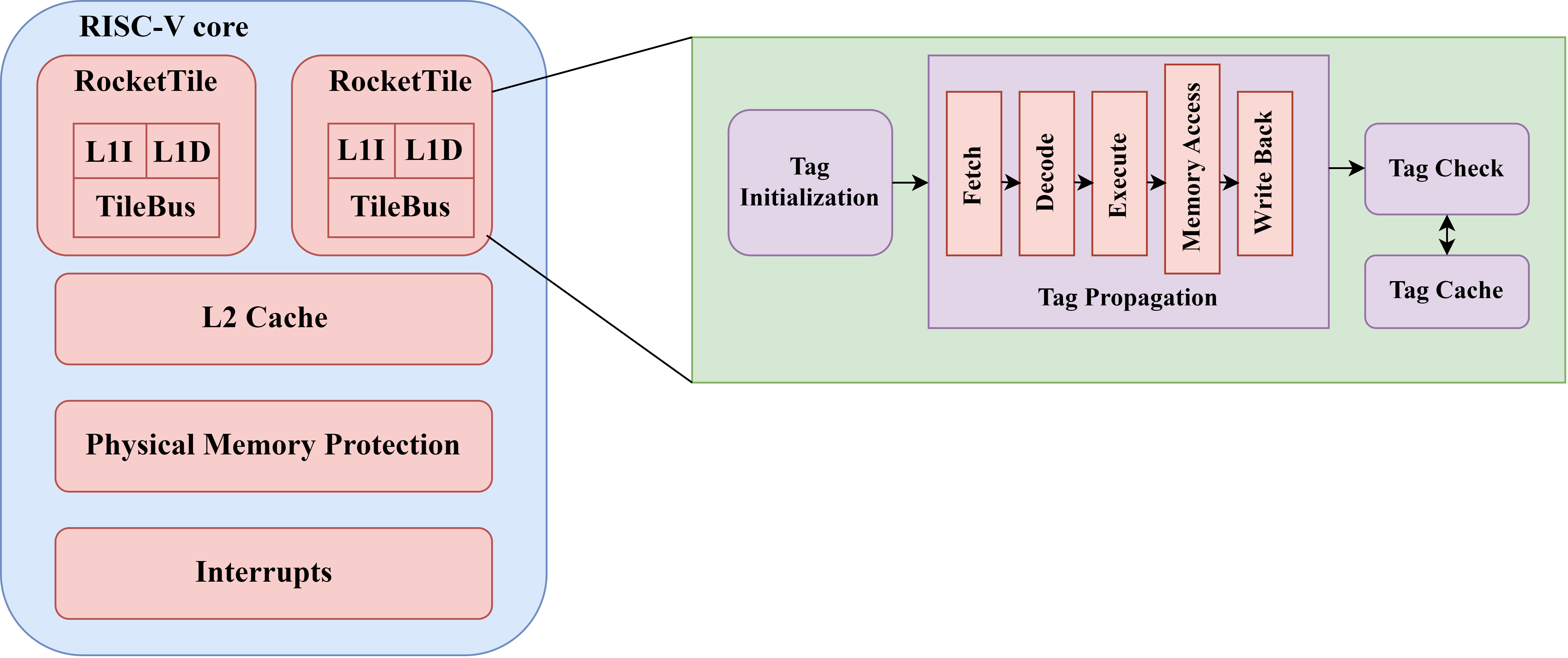}
\caption{Tag module.\label{fig7}}
\end{figure}

The ISA-level IFT model is integrated at the microarchitecture level and has introduced secure instructions, specifically, load and store. The proposed technique is demonstrated on the RISC-V processor’s datapath and control logic.
The integrated gate-level IFT uses the data-level logic of only the security-critical modules. 
Both models interact using the tilelink interface, as shown in Figures~\ref{fig6} and~\ref{fig7},  with the expanded core integration of the tag module inside the core. 

As shown in Figure~\ref{fig7}, the ISA-level tag module in the RISC-V core is updated as the signal propagates through the datapath at the coarse-grained level. The processor datapath cycle is modified with the tag initialization, tag propagation and tag check processes for the secure load and store instructions. The first process of the tag initialization module adds a 1-bit tag to the new instruction and moves to the next process of tag propagation. The tag propagation module integrates the tag propagation through the instruction cycle of fetch, decode, execute, memory access and write back for the secure load and store instructions. Finally, the tag check module checks the tag bit associated with the tag cache with security policies to raise an exception. 

\subsubsection{Security Policies}

Potentially malicious input/output channels for a system are identified via security policies~\cite{ref-p45,ref-p46}. Focusing on memory corruption, the security policies are associated with the load/store instructions and the return addresses. All user-supplied data are marked as malicious since they use the load/store instructions with source and destination addresses. The program counter value is also tracked as the return address and needs to be protected from modification. The tag check rules are applied to the secure custom instructions with tag bits, and an exception is raised if the tag values mismatch to mitigate the attack. 

If a conditional branch instruction is executed, the tag check module checks for the address using the tag bits and permits the execution only if the tag bits match with the table or terminate the call, thus providing a coarse granularity model with better precision. The security policies for protecting the return address assign checking rules for tracking all procedure calls, along with user input data.

\subsection{Gate-Level Model}
Gate-level model components track security-sensitive paths using fine-grained information flow at the logic blocks by performing a composition of augmented logic blocks~\cite{ref-p47}. In the fine-grained IFT, implicit, explicit and covert information flow is tracked at the data level. In an integrated CF-IFT scheme, the gate-level design focuses on a specific security critical module and its corresponding datapath to minimize the security overhead and performance, as compared to only gate-level IFT models. Special critical modules, such as crypto engines and accelerators, that share secret keys or security-critical information are considered the logic blocks to which gate-level IFT is applied. 

Without any hardware design modification, this approach is integrated with the RISC-V core as a separate module to perform information policies for user-instantiated critical modules. Each input and output of the entire module is tainted, and a shadow logic library for the tainted gates is implemented. Security policies are matched with the tainted information for the module under test. The tainted inputs for the gates are determined by detecting the arbitrary changes in the inputs that affect the output, leading to data leakage and faults and triggering hidden functionalities.

\subsubsection{Shadow Logic for Tainted Gates  }

\textls[-20]{All data bits propagating from the input to the output are marked as tainted if the tainted input influences the output. Considering the example in Figure~\ref{fig2}, we have two OR gates and one AND gate, with a and b as inputs and o as the output. In an OR gate, with reference to the truth table for the two inputs, additional tainted inputs are added, such that when the inputs are toggled and output is affected by the tainted input,
it is considered malicious and marked to be tracked through the whole information flow. Based on the minterms, the shadow logic is formed for the OR gate, and similarly, this is conducted for the AND gate.}

The shadow gate library is formed for all basic gates and compared with the main logic to mark the tainted output bits. The information flow policy checking module is used to track the flow based on the security policies implemented. Precision logic is indicated based on the tainted information. If the security policy is violated, the output flowing from the input is considered spurious with information leakage.

\subsubsection{Information Flow Policy Checking}

The security policies for the logic gates and the associated tainted bits are implemented for all the gates in comparison with the shadow logic library~\cite{ref-p48}. Initially, the module to be tracked is selected by the user. The module or the application is then converted to a gate-level netlist to track the information flow of individual data and is separated into sub-modules.
The gate-level IFT module obtains the information and security policies from the shadow logic and the IFT policy checking rules unit to determine the tainted inputs, the exact location of data leakage and spurious inputs. The gate-level IFT unit tracks these untrusted information paths and raises an exception if there is a change in the output and information flow, resulting in malicious data detection. 

\begin{figure}[H]

\includegraphics[width=9.5 cm]{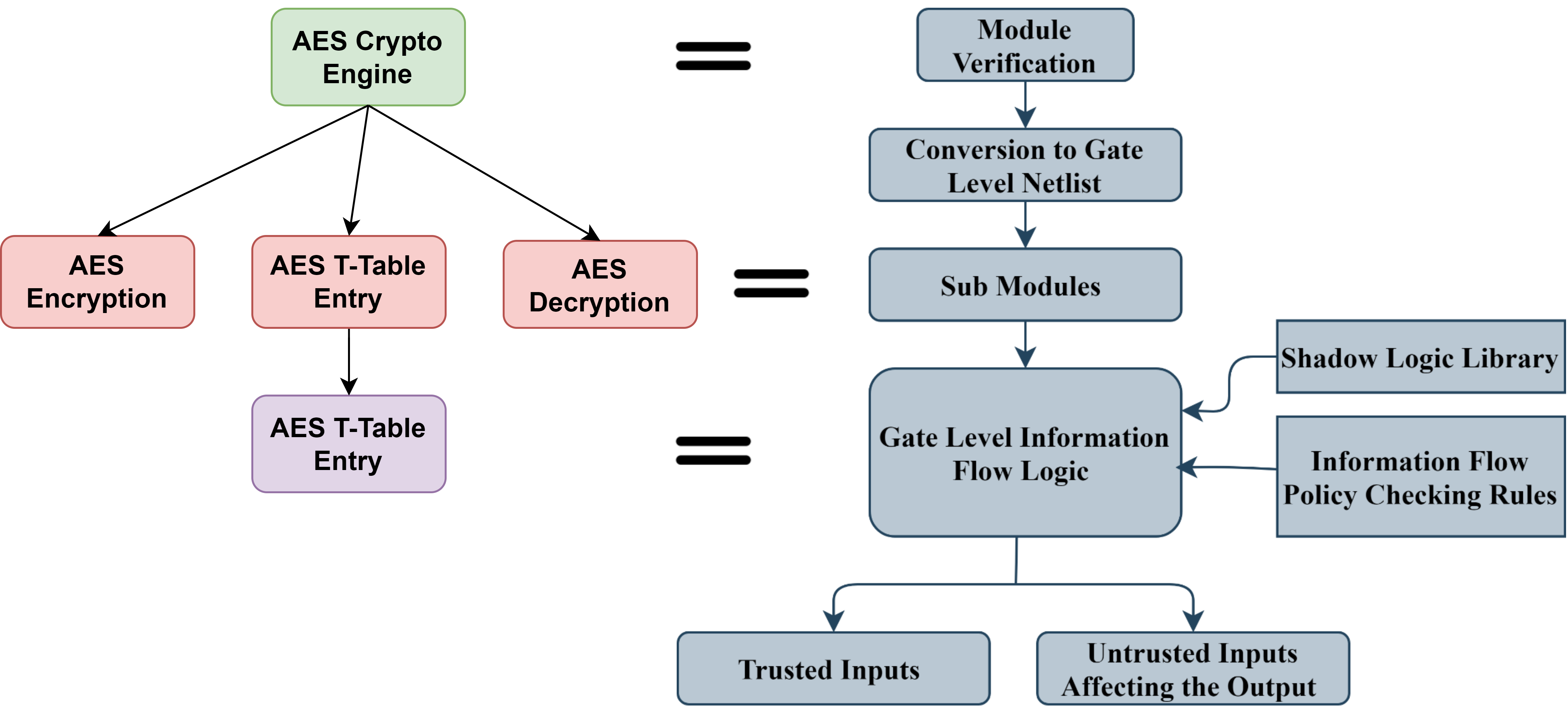}

\caption{Gate-level IFT model flow. \label{fig8}}
\end{figure}

\textls[-20]{Figure~\ref{fig8} demonstrates the high-level automation flow of gate-level IFT integration to the security-critical datapaths to the cryptographic cores. The cryptographic cores are complex and may have multiple rounds performing computationally extensive operations. The core is evaluated on the security properties and selects the submodules that are critical to perform runtime execution analysis. As shown in Figure~\ref{fig8}, the Advanced Encryption Standard (AES) module is divided into the submodules of the encryption and decryption cores that use T-tables for the substitution of bytes. T-tables are vulnerable to cache attacks that can maliciously update the substitution and/or leak information. The T-table sub-module is selected for fine-grained gate-level IFT integration to monitor signals using shadow logic. The gate-level IFT unit tracks the untrusted information paths, raises an exception if there is a change in output and detects malicious data propagation during runtime.}

 
 This approach is efficient with no hardware modification and integrated along with architecture-level IFT, which enables an accurate and performance-efficient model. Depending on the application or micro-architectures added to the system, gate-level IFT is performed on any critically sensitive modules. For example, in crypto engines, the keys in the AES encryption module can be tracked using gate-level IFT to check integrity and data leakage from the system.
 
\begin{figure}[H]

\includegraphics[width=6.5 cm]{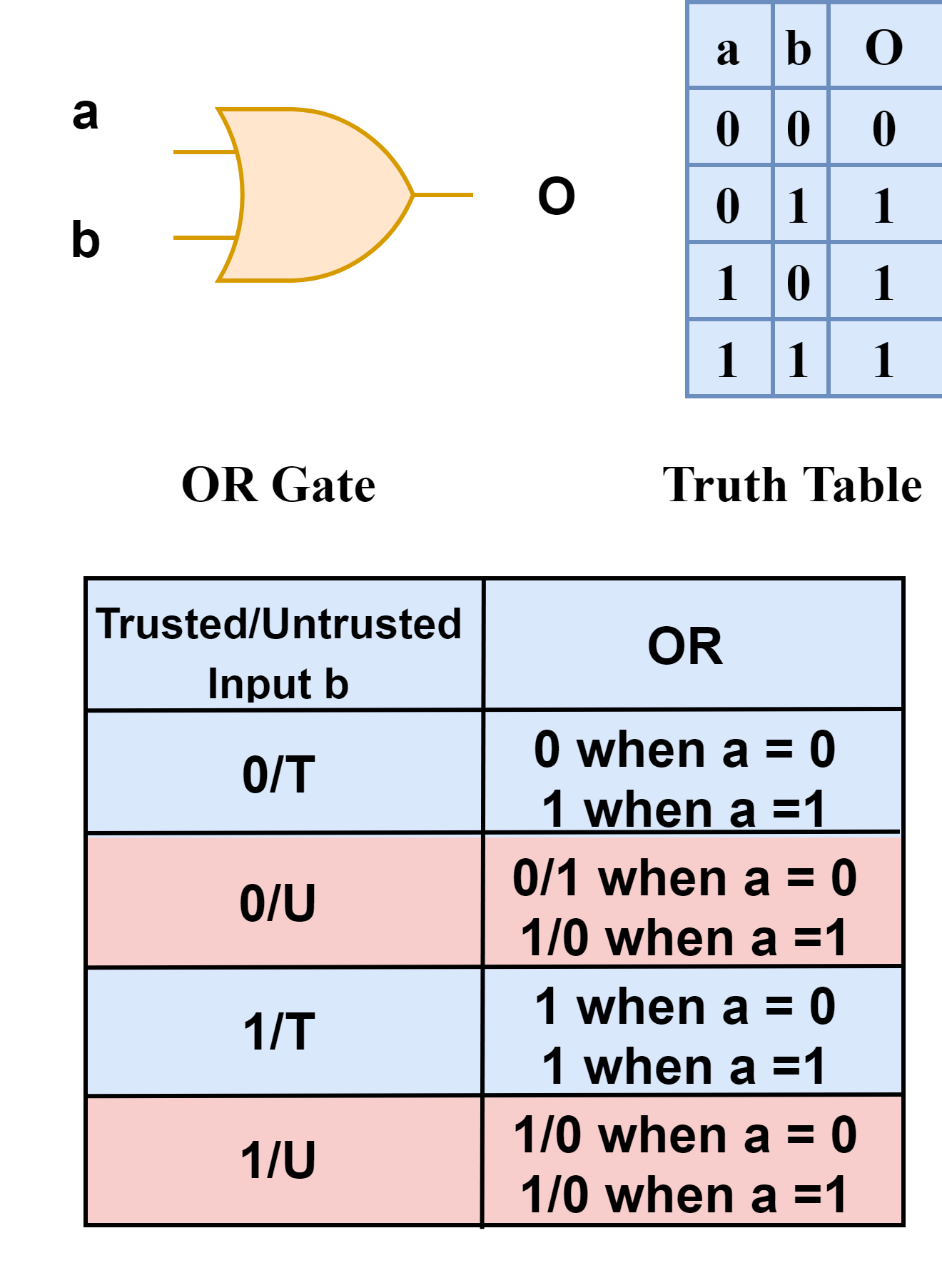}
\caption{OR gate shadow logic truth table. \label{fig9}}
\end{figure}

 Figure~\ref{fig9} shows the simplified shadow logic for a two-input OR gate. In this example, input b is considered untrusted, and the shadow logic truth table is highlighted to show the tainted input affecting the output. If a = 0 and b is untrusted, then the output (o) = 0/1, and when a = 1 and b is untrusted, then the output (o) = 1/0, which proves that the output is affected via the input b. Similarly, the shadow logic library holds the truth table for all gates, the tainted data are tracked using this logic and the exact location that leads to data leakage is found.

\subsection{Simulation Model}

 Spike is a RISC-V ISA simulator, with the ISA specification integration, that acts as a reference model for RISC-V. This work extends the simulator capabilities and enables new security extensions to detect runtime attacks in the RISC-V simulation model. The capability enriches the software simulation model to correlate with the hardware model to monitor runtime IFT. This is, to the best of the author's knowledge, the first effort to integrate security features at the simulation level and can result in the security integration of EDA tools.
 
\textls[-20]{The goal of compiler/assembler modification is to extend the ISA for several instruction set extensions. The Spike ISA simulator is used to test the modification of ISA at the software level and is divided into different modules. Each module is responsible for simulating a block of architecture, and adding new blocks in the simulator yields new security extensions in the model. The initial step is to declare the instruction in the instruction format to generate matching and masking address values for the new instruction. The following step is to describe the instruction length, number of operands and functionality, as the hardware model along with the corresponding address and master address are introduced.}

\begin{figure}[H]

\includegraphics[width=12 cm]{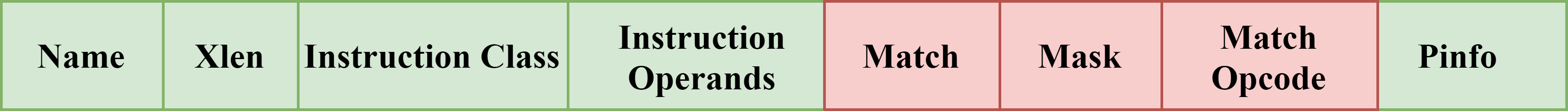}
\caption{Field format to add new instructions in RISC-V.\label{fig10}}
\end{figure}

Figure~\ref{fig10} shows the field format for the RISC-V architecture to add the new instructions in the RISC-V opcode structure. The highlighted red fields are modified to accommodate the new instructions in the toolchain. The check tag and store tag functions are implemented to execute the tag condition for the return address along with security policies to determine the tagged flow.

To simulate the newly added instructions in the Spike simulator, functions are added to the tag module to check and store tag bits. For the IFT simulation module,
a buffer overflow attack in which the user input data are untrusted is implemented and demonstrates that the secure IFT model can detect the modified return address and mitigate the attack. The location of the stored return address on the stack is retrieved by causing the buffer overflow attack. The toolchain support provides the flexibility to add security policies and develop test beds to carry out security analysis for RISC-V ISA with minimal design overhead and better precision logic.
\section{Experimental Results}\label{sec5}

\subsection{Architectural Model}

\begin{figure}[H]
\includegraphics[width=10.5 cm]{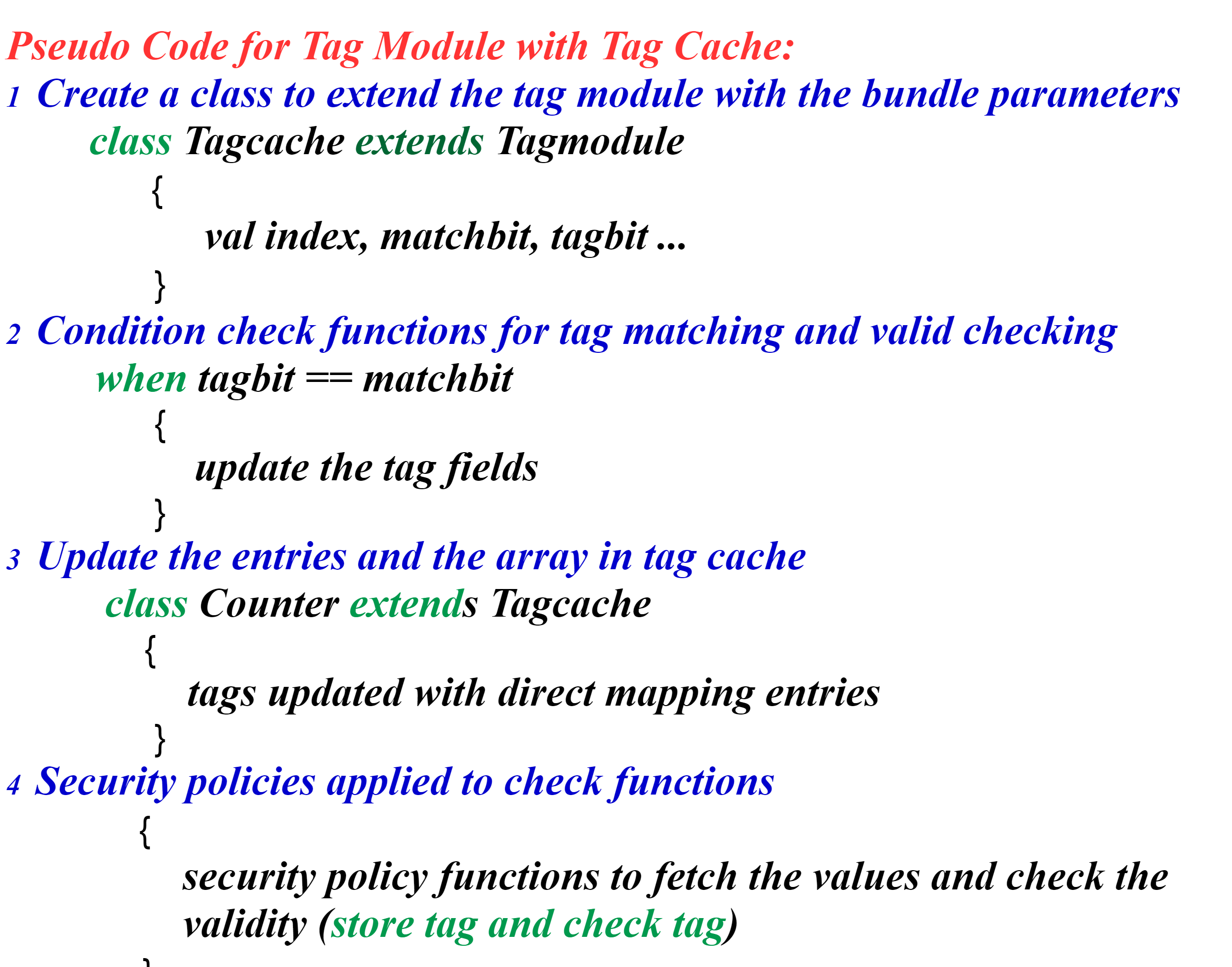}
\caption{Pseudo code for tag module with tag cache.\label{fig11}}
\end{figure}

In the tag module, the tags labeled for the return addresses and data are processed, and the information is stored in the tag cache. Similar to the load/store instructions, the security-enhanced tag-based load/store instructions are used for adding tags to only the return addresses and the user-defined data. Figure~\ref{fig11} shows the pseudocode for the tag module along with the tag cache, where a class is created for the tag cache with various tag parameters. Check functions are implemented for tag matching, which includes a counter to maintain the tag array for the tag cache and is updated based on entries. Security policy functions are written to fetch the tags and match the tag bits for validation.
\vspace{-6pt}

\textls[-20]{The Xilinx Artix-7 Field Programmable Gate Array (FPGA) board is used to prototype and run the modified RISC-V rocket chip using Vivado. A user-defined application is modified to run a buffer overflow attack to test the architecture. The LookUp Table and Flip Flops for the proposed IFT model are compared with the Cryptographic Return Address Stack (CRAS)-based countermeasures~\cite{ref-p16}, as shown in Table \ref{tab1}. The proposed model uses 10528 LUTs and 7113 Flip Flops to implement the proposed IFT model in the RISC-V architecture, which is fewer resources than the CRAS model. The CRAS model further uses a separate stack for storing the return addresses to eliminate data leakage from shared memory. The result shows that the overall increase in the usage of LookUp Table (LUT) resources does not exceed 1\%. The overall FPGA resources, which are used in addition to the application, are used as the tags and a copy of inputs processed in a parallel manner. The tag module utilizes the distributed LUTs of the application and does not impact the processor’s area overhead.}

\begin{table}[H] 
\caption{Resource utilization of the IFT module.\label{tab1}}
\newcolumntype{C}{>{\centering\arraybackslash}X}
\begin{tabularx}{\textwidth}{cCC}
\toprule
\textbf{Architecture }	& \textbf{LUT}	& \textbf{FF}\\
\midrule
Proposed tag-based IFT	& 10,528			& 7113\\
\midrule
Cryptographic Return Address Stack (CRAS)		& 11,468			& 6130 \\
\bottomrule
\end{tabularx}
\end{table}

\subsection{Gate-Level Model}
The security properties of the gate-level IFT and verification are analyzed on ISCAS-85 Benchmark circuits using the critical datapaths. 

\subsubsection*{Netlist and Submodules }

The combinational circuit c432 interrupt controller is considered to illustrate the gate-level IFT implemented in the hardware. The circuit consists of 36 inputs and 7 outputs with a total of 160 gates in which each channel is enabled on priority based on the bus requests. The initial circuit is converted to a gate-level netlist using the implemented algorithm and is verified as the module to be tracked. The module is classified into submodules based on the information and critical data. Finally, the gate-level IFT logic is used to compare the submodule to be tracked with the shadow logic library. The security policies are applied to track the gates in which the output is affected by the input. This gives a set of trusted input gates and untrusted input gates that need to be tracked. 

The gate-level IFT model tracks the untrusted gates and detects information leakage by analyzing the data of all untrusted gate inputs. A variation in the data flow is observed and raises an interrupt, indicating a malicious information flow that is hidden in the ISA abstraction. This model provides fine precision logic to accurately predict untrusted data. Implemented as a separate automated model, this provides less area overhead when compared to other existing GLIFT models, which increases the area by 70\% by doubling the gates for shadow logic.  

\begin{figure}[H]

\includegraphics[width=10.5 cm]{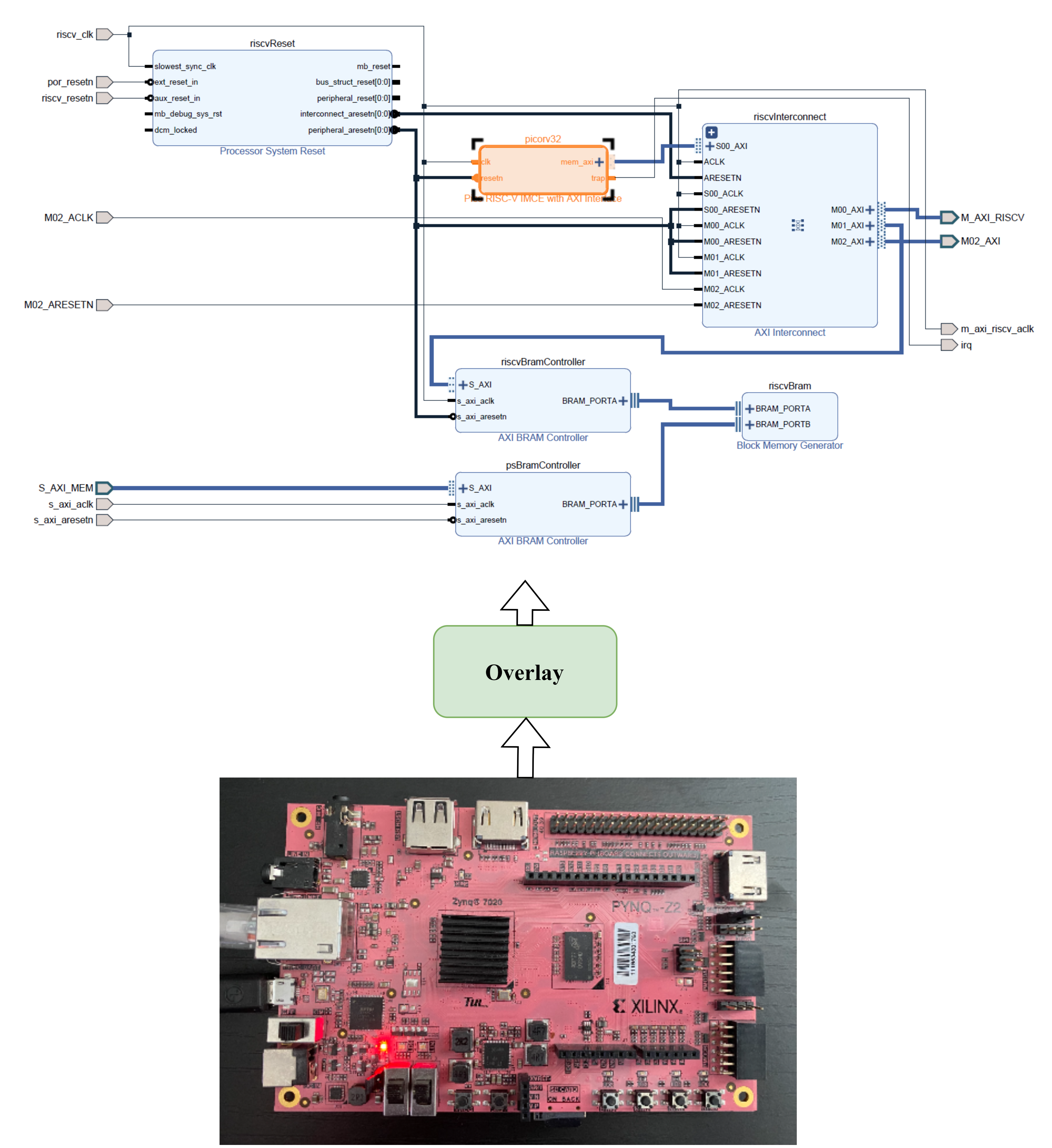}
\caption{PYNQ flow with the RISC-V IP and AXI Interface.\label{fig12}}
\end{figure}

The Xilinx PYNQ FPGA board is used to demonstrate the gate-level IFT model with the c432 ISCAS-85 benchmark circuits. As the RISC-V specification is highly flexible, different soft processor variants can be used as a tool to study the threat model. Building flexible Inter and Intra-ISA variations for comparison with the RISC-V family helps in analyzing the security extensions. RISC-V IP is implemented using the Vivado tool for the PYNQ board, supporting the overlay. The overlay acts as a bridge between the designed IP and the FPGA. Figure~\ref{fig12}  shows the PYNQ flow with the RISC-V IP and  Advanced Extensible Interface (AXI) designed for the PYNQ board using the overlay. GLIFT is integrated with the IP to track the datapath of the security-critical modules. 

\begin{figure}[H]

\includegraphics[width=10.5 cm]{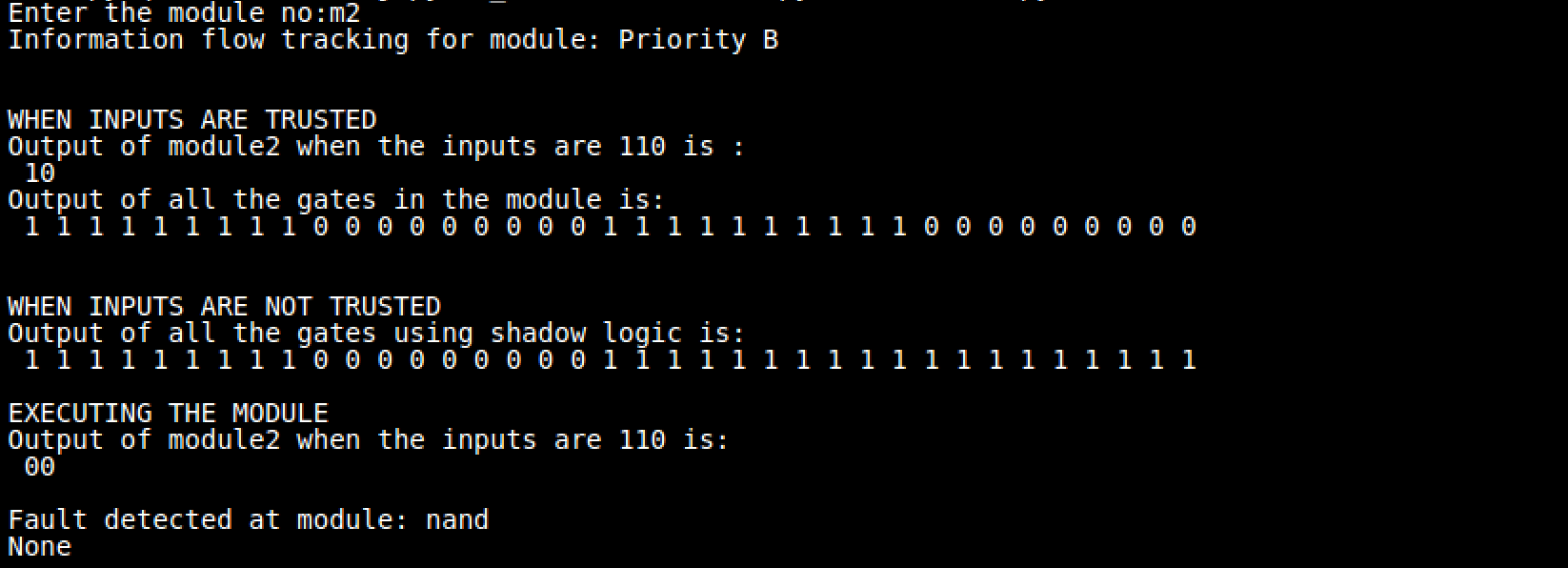}
\caption{Gate-level IFT detecting the untrusted gate (NAND) for the submodule priority encoder B in the c432 circuit.\label{fig13}}
\end{figure}

Figure~\ref{fig13}  shows the gate-level IFT detecting the untrusted gate in the submodule m2 for the c432 circuit. The circuit is divided into submodules, from which we focus on one of the modules, that is, module 2, implementing priority encoder B, as shown in Figure~\ref{fig13}  line 2. Priority encoder B is selected to be tracked with untrusted inputs, and the model detects the fault at the NAND gate based on the shadow logic libraries and the sequence of inputs. It classifies the gates based on the shadow gate libraries and points out the untrusted gate. Similarly, with different sets of input, the model detects the untrusted gates and tracks the gates further during runtime to detect the leakage of critical data and the modification of data inputs and raises a security function call if there is a detection of malicious data inputs. 

\textls[-25]{Table~\ref{tab2} shows the execution of the shadow logic to detect the untrusted gates in the circuit of different benchmarks. Individual security critical modules that are vulnerable to data leakage through input datapath can be tracked to identify the location of the untrusted inputs based on the proposed model, which improves the execution time, as the other paths do not require shadow logic integration, leading to a precise model without design modification or scaling.}

\begin{table}[H] 
\tablesize{\footnotesize}
\caption{
The execution time of ISCAS 85, 89, as well as the EPFL benchmarks.\label{tab2}}
\newcolumntype{C}{>{\centering\arraybackslash}X}
\begin{tabularx}{\textwidth}{cccCC}
\toprule
\textbf{Benchmark }	& \textbf{No. of Inputs}	& \textbf{No. of Outputs}   & \textbf{No. of Gates}   & \textbf{Shadow Logic Execution (s)}\\
\midrule
C432	& 36		& 7   & 160    & 186\\
\midrule
S398	& 3		& 6   & 119  & 177 \\
\midrule
Adder   & 256	& 129   & 2162  & 556 \\
\midrule
Alu + crtl    & 7	& 26   & 306  & 373 \\
\midrule
Memory-crtl (submodule) & 1204	& 1231  & 8956  & 984 \\
\bottomrule
\end{tabularx}
\end{table}

For the GLIFT, the resource utilization shown in Table~\ref{tab3} indicates that the proposed RISC-V-based GLIFT uses only 4506 LookUp Tables, 4753 Flip Flops, 188 LUTRAMs and 16 BBRAMs, as compared to the overall IFT-integrated RISC-V SoC. The overall coverage of the IFT-integrated datapaths and control paths is optimized, as the security-critical paths are identified for the fine-grained IFT, and the other paths are controlled via the microarchitectural-level IFT. This integrated IFT scheme reduces area overhead without penalizing the detailed IFT control over the security datapaths.

\begin{table}[H] 
\tablesize{\footnotesize}
\caption{Resource utilization of GLIFT module.\label{tab3}}

\newcolumntype{C}{>{\centering\arraybackslash}X}
\begin{tabularx}{\textwidth}{CCCCC}
\toprule
\textbf{Architecture}	& \textbf{LUT}	& \textbf{FF}   & \textbf{LUTRAM}   & \textbf{BBRAM}\\
\midrule
SoC with proposed IFT integration	& 53,200		& 106,400   & 17,400    & 140\\
\midrule
Proposed IFT-only overhead & 4506		& 4753   & 188   & 16 \\
\midrule
Percentage overhead of IFT integration & 8.5\%		& 4.5\%   & 1.1\%  & 11\%\\
\bottomrule
\end{tabularx}

\end{table}

\begin{figure}[H]

\includegraphics[width=10.5 cm]{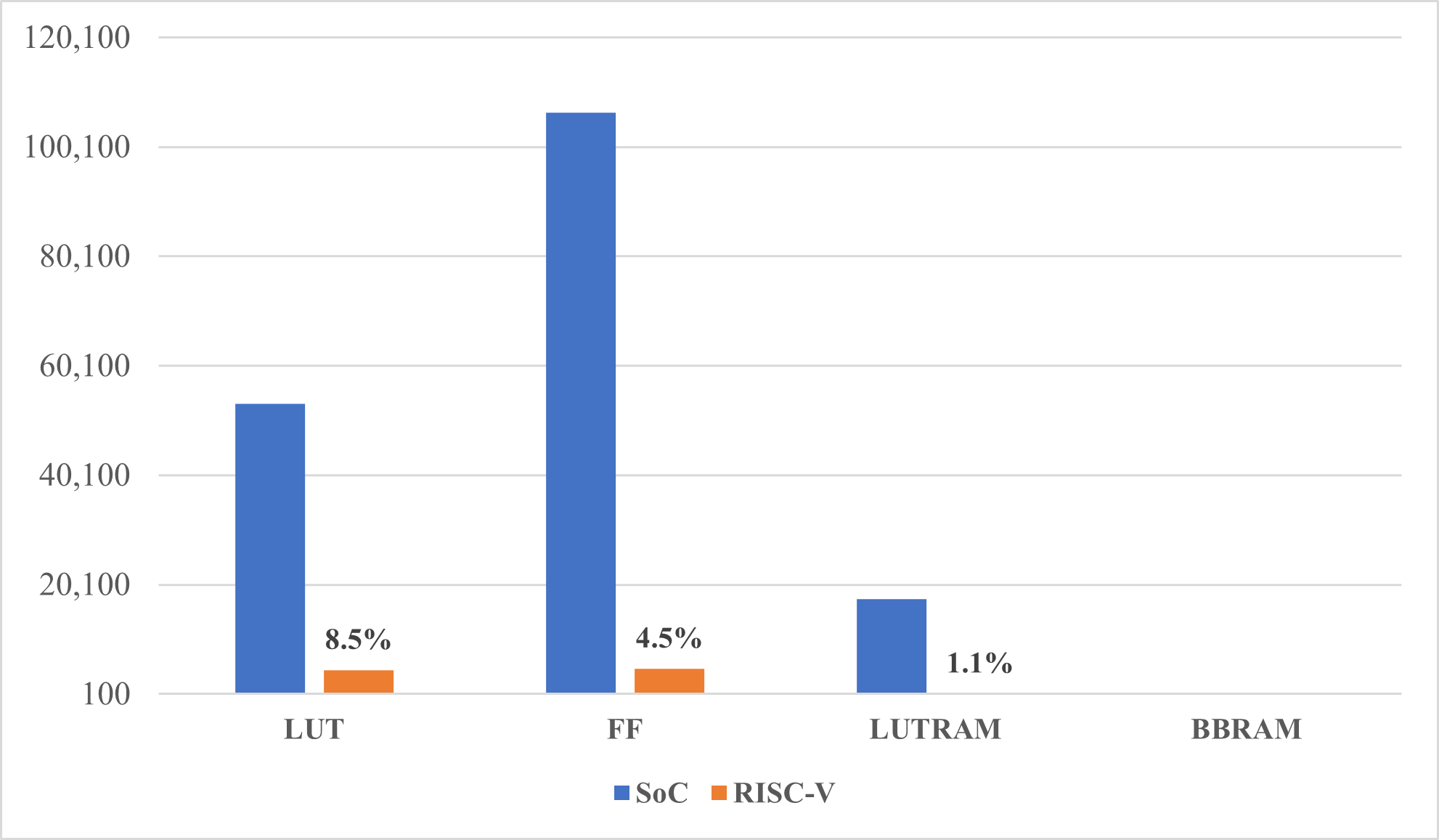}
\caption{Relative resource utilization.\label{fig14}}
\end{figure}

Figure~\ref{fig14}  shows the relative utilization percentage graph with the SoC, where only 8.5\% of the LUT is utilized for the RISC-V IP.

\subsection{Simulation Model}
 
\textls[-20]{The IFT framework is tested on the Spike simulator by implementing a buffer overflow attack. The new instructions are used with the stack operations where the return address is saved. Tag bits are assigned to the return address and the data to be stored on the stack. If there is a mismatch, an exception is raised, and the program execution is stopped by mitigating the buffer overflow attack. The newly added SDTCHECK is used to assign a tag bit for the return address, and the LDTCHECK is used to check the tag corresponding to the return address in the tag cache. This framework protects the stack by using customized instructions where the new return address compromised by an adversary is not loaded on the stack.}

\begin{figure}[H]

\includegraphics[width=10.5 cm]{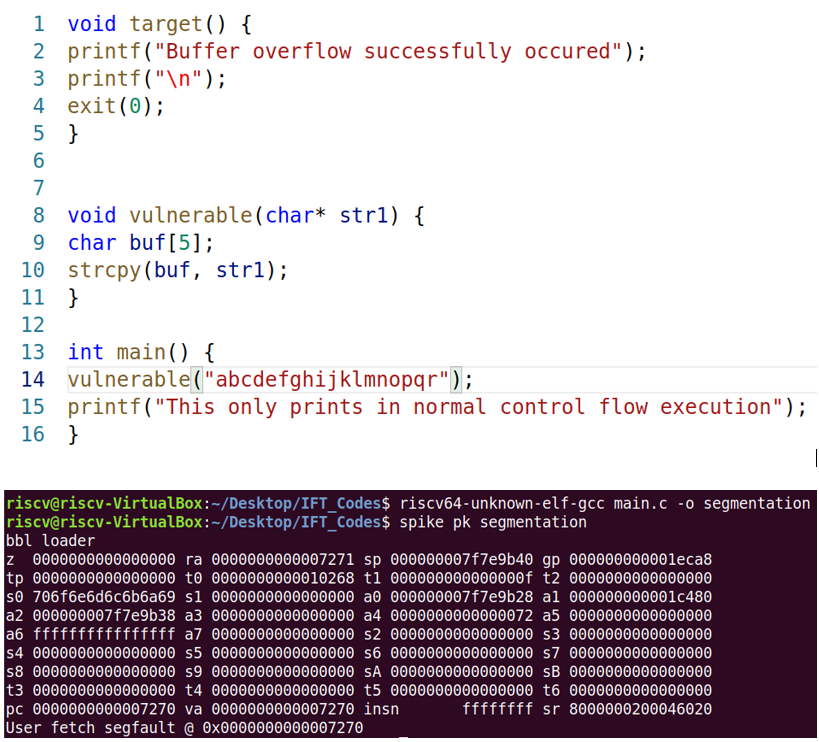}
\caption{Buffer overflow attack in C with a segmentation fault.\label{fig15}}
\end{figure}

Figure~\ref{fig15}  shows a buffer overflow C code with a segmentation fault, in which the vulnerable function has an argument str1 passed to the buffer of size 5 using string copy. This will cause a segmentation fault if the allocated memory address is exceeded, but if bypassed, it may access restricted locations and an adversary can gain access to the return address. This proves that even though the segmentation fault exception exits the program execution and prevents the program from accessing the protected memory location stored on the stack, it can still cause leakage of critically important data.

\begin{figure}[H]

\includegraphics[width=10.5 cm]{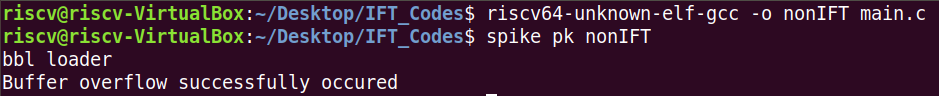}
\caption{Buffer overflow attack in RISC-V architecture without IFT module.\label{fig16}}
\end{figure}

\begin{figure}[H]

\includegraphics[width=10.5 cm]{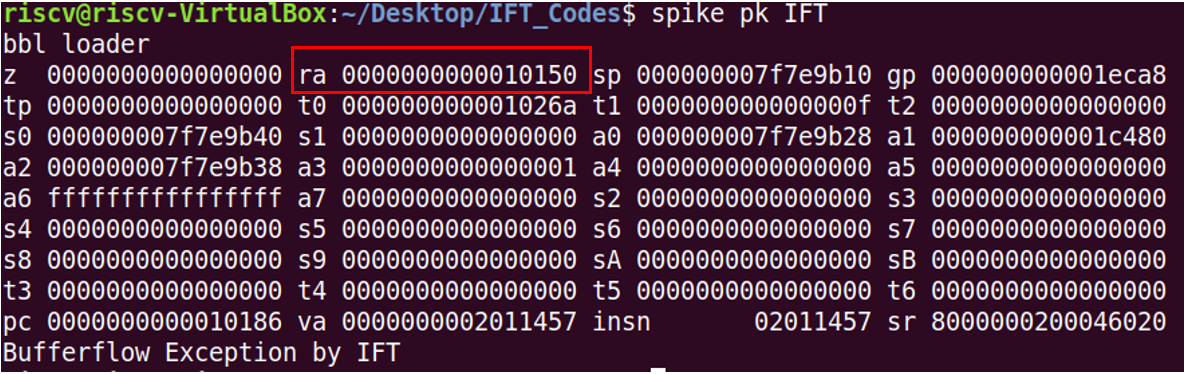}
\caption{IFT module implemented eliminating the attack via exception.\label{fig17}}
\end{figure}

When the segmentation fault is bypassed, the program executes the conventional RISC-V architecture by successfully executing the buffer overflow attack by modifying the return address. Such a scenario will cause a memory attack and can be mitigated using the proposed IFT model, that is, once applied, it results in an exception to terminate the process. The model identifies the modification of the return address and stops the program execution by raising an exception to eliminate the buffer overflow attack, as shown in Figure~\ref{fig16}. Figure~\ref{fig16} shows the buffer overflow attack executed successfully without the IFT model on RISC-V, and Figure~\ref{fig17} shows the IFT module tracking the return address and eliminating the attack by raising an exception when the return address is modified.

\section{Security Analysis}\label{sec6}

In this section, the security extensions provided by the proposed framework to eliminate memory-based attacks are discussed:
\begin{itemize}
    \item The RISC-V model is vulnerable to leaks such as buffer overflows, return address attacks, fault injections, etc. An adversary can modify the bit flips or the return address of the application, leading to compromised data. The data flow of the security-critical data is analyzed for accuracy via the tag module, providing runtime protection for data integrity. The security-enhanced RISC-V model thwarts these vulnerabilities via the multi-granularity Information Flow Tracking to enable early capture and mitigation, that is, before the malicious instructions of data can impact the outputs or control flow.
    \item The control-flow hijack of the security-critical modules leaking data or modifying the IP cores is an important security verification property that is handled by the gate-level IFT, which restricts the integration to only the critical paths, enabling minimal overhead in contrast to designs that support full coverage, which add more design overhead. It provides accuracy with all information flows along all timing channels. 
    \item The simulation model with toolchain extensions supporting custom ISA is flexible to add new security policies for security analysis. It supports the architectural design and verification of security extensions to the RISC-V processor. The simulation model prototypes for accelerating the security verification from an architectural aspect to compiler-specific verification. Thus, providing a simulation model makes the verification process faster and enables fast integration.
    \item The proposed model protects the system against leakage models by using gate-level logic and maintains the data-flow integrity by eliminating memory corruption attacks with accuracy. Thus, the model uses the best of the existing schemes and provides security at different levels of abstraction.
    \item The proposed design is a flexible architecture with extensibility that can support the Common Evaluation Platform (CEP) reference model to protect the system during runtime. This enables the integration of research to further improve state-of-the-art security verification for the RISC-V platform, thus supporting hardware security research, such as side-channel analysis. 

\end{itemize}
\section{Limitations and Future Work}\label{sec7}

\textls[-20]{The current prototype of the IFT model focuses on modifying the datapath at a microarchitectural level. It provides a multi-granularity model that demonstrates the flexibility of using different soft processor RISC-V architecture variants to study threat models and analyze security extensions. The results obtained so far are limited to simple applications, with an emphasis on area overhead and resource utilization in the implemented design. The model achieves security mechanisms, such as stack protection, with information leakage prevention.}

In future work, real-time applications and benchmarks will be incorporated to evaluate the correctness, efficiency and effectiveness of the implemented model. The system aims to identify untrusted sources using tag mechanisms and leakage models in specific units through the use of shadow logic. However, there are tradeoffs to consider in this model:

Tag Table: The two new instructions used for the tag mechanism rely on a tag table, which can be tampered with using different software attacks. To mitigate this, protection techniques can be implemented to secure the tag table, and an on-demand allocation mechanism for tag entry can be employed.

Heap Exploit: The attack scenario in this model focuses only on stack protection, whereas heap overflows in the dynamically allocated memory can be exploited to overwrite data in the heap. Further research and enhancements are needed to address this limitation.

Verification Mechanism: Runtime IFT mechanisms can monitor realistic information flow behaviors but come with higher area and performance overheads. This is because it is an architectural approach with different levels of abstraction. Balancing the need for accurate monitoring with the associated overhead is an ongoing challenge.

Overtainted Tag Propagation: The tag propagation in this model may result in high false positives, as tags can pollute other data during runtime. For example, if a tag is propagated from a general-purpose register through instructions involving memory locations, the data associated with the initial tag may be considered unsafe, leading to unnecessary exceptions being raised. This issue can be mitigated by assigning tag memory mapping and implementing restricted control-flow security policies.

These tradeoffs highlight areas for further research and improvements to the IFT model, ensuring a more robust and effective approach to analyzing security extensions and mitigating potential threats.

\section{Conclusions}

\textls[-20]{This study proposes a new multi-granularity-based IFT model that utilizes a tagging mechanism for data-flow integrity and an optimized shadow logic for the leakage model. The modified RISC-V core with the security extensions of IFT provides system protection by tracking the data flow from untrusted channels. Furthermore, the hardware-architecture-specific extensions are translated to a compiler-specific simulation model with minimal design overhead and verified with a buffer overflow attack model. The hardware design shows less than 1\% of area overhead with better precision logic. In future work, the security policies will be further enhanced to track the side-channel vulnerabilities for security-critical data.} 


\vspace{6pt} 

\authorcontributions{
Conceptualization, G.S.N. and F.S.; Methodology, G.S.N., D.V.A.; Software, G.S.N., B.T.; Validation, G.S.N.,D.V.A., B.T. and F.S.S.; Formal analysis, G.S.N. and F.S; Investigation,G.S.N.; Resources,G.S.N. and F.S.; Writing—original
draft preparation, G.S.N.; Writing—review and editing, F.S. and G.S.N.; Visualization, G.S.N.; Supervision, F.S.;
All authors have read and agreed to the published version of the manuscript.}

\funding{This research was funded by NSF (grant numbers: 1814420, 1819694 and 1819687).}

\dataavailability{No new data were created or analyzed in this study. Data sharing is not applicable to this article.}

\conflictsofinterest{The authors declare no conflicts of interest.}

\begin{adjustwidth}{-\extralength}{0cm}

\reftitle{References}

\PublishersNote{}
\end{adjustwidth}
\end{document}